\documentclass[aps,prd,preprint,tightenlines,groupedaddress,amsmath,amssymb,superscriptaddress]{revtex4}  
\usepackage{graphicx}
\renewcommand{\vec}[1]{\mathbf{#1}}
\begin{document}
\title{Measuring the 3D Clustering of Undetected Galaxies Through Cross Correlation of their Cumulative Flux Fluctuations from Multiple Spectral Lines}

\author{Eli Visbal}
\email[]{evisbal@fas.harvard.edu}
\affiliation{Jefferson Laboratory of Physics, Harvard University, Cambridge, MA 02138}
\affiliation{Harvard-Smithsonian CfA, 60 Garden Street, Cambridge, MA 02138}

\author{Abraham Loeb}
\email[]{aloeb@cfa.harvard.edu}
\affiliation{Harvard-Smithsonian CfA, 60 Garden Street, Cambridge, MA 02138}

\date{\today}

\begin{abstract}
We discuss a method for detecting the emission from high redshift
galaxies by cross correlating flux fluctuations from multiple spectral
lines.  If one can fit and subtract away the continuum emission with a
smooth function of frequency, the remaining signal contains
fluctuations of flux with frequency and angle from line emitting
galaxies.  Over a particular small range of observed frequencies,
these fluctuations will originate from sources corresponding to a
series of different redshifts, one for each emission line.  It is
possible to statistically isolate the fluctuations at a particular
redshift by cross correlating emission originating from the same
redshift, but in different emission lines.  This technique will allow
detection of clustering fluctuations from the faintest galaxies which
individually cannot be detected, but which contribute substantially to
the total signal due to their large numbers.  We describe these
fluctuations quantitatively through the line cross power spectrum.  As
an example of a particular application of this technique, we calculate
the signal-to-noise ratio for a measurement of the cross power
spectrum of the OI(63 $\mu$m) and OIII(52 $\mu$m) fine structure lines
with the proposed Space Infrared Telescope for Cosmology and
Astrophysics (SPICA).  We find that the cross power spectrum can be
measured beyond a redshift of $z=8$.  Such observations could
constrain the evolution of the metallicity, bias, and duty cycle of
faint galaxies at high redshifts and may also be sensitive to the
reionization history through its effect on the minimum mass of
galaxies.  As another example of this technique, we calculate the
signal-to-noise ratio for the cross power spectrum of CO line emission
measured with a large ground based telescope like the Cornell Caltech
Atacama Telescope (CCAT) and 21-cm radiation originating from hydrogen
in galaxies after reionization with an interferometer similar in scale
to the Murchison Widefield Array (MWA), but optimized for
post-reionization redshifts.
\end{abstract}

\maketitle

\section{Introduction}
Atoms and molecules in the interstellar medium of galaxies produce
line emission at particular rest frame wavelengths \cite{binney}.  For
galaxies at cosmological distances, this line emission is redshifted
by a factor of $(1+z)$ due to the expansion of the universe.  Ignoring
peculiar velocities, the redshift of a galaxy corresponds to its
distance from the observer along the line of sight.  Thus, if the
redshift and angular position of many galaxies are measured, a 3D map
of their distribution can be constructed.  Such 3D maps contain a
wealth of information about galaxies and the underlying cosmic density
field which they trace.  Examples of recent galaxy surveys include the
Sloan Digital Sky Survey \cite{SDSSref}, the 2dF Galaxy Redshift
Survey \cite{2dFref}, the DEEP2 Redshift Survey \cite{DEEP2ref}, and
the VIMOS-VLT Deep Survey \cite{VIMOSref}.

In this work, we discuss a different way of mapping the large scale
structure of the universe, namely by measuring the fluctuations in
line emission from galaxies, including also those too faint to be
individually detected.  If the line emission fluctuations in observed
frequency and angle are associated with a particular line, they can be
translated into a 3D map.  Even though individual faint sources cannot
be distinguished from noise, the cumulative emission from many such
sources contributes substantially to the fluctuations over large
scales.  These fluctuations will trace the large scale structure of
the universe and can be analyzed statistically.  By contrast, a
traditional galaxy survey only contains the positions of galaxies
detected at high significance, and discards information from fainter
sources.

Similar observations are being planned with the 21cm line of neutral
hydrogen using instruments such as MWA \cite{MWAref}, LOFAR
\cite{LOFARref}, PAPER \cite{PAPERref}, 21CMA \cite{21CMAref}, and SKA
\cite{SKAref}.  These experiments will obtain angle and frequency
information for large areas on the sky and will produce 3D maps of
the neutral hydrogen throughout the universe.  The raw signal will
contain both bright sources of foreground emission (the largest being
synchrotron emission from our galaxy), as well as the cosmological
21cm line signal.  Since the cosmological signal varies rapidly with
frequency, it is expected that the foregrounds, which are smooth in
the frequency direction, can be subtracted off
\cite{2005ApJ...619..678M,Wang:2005zj,2006ApJ...653..815M,2009MNRAS.398..401L}.
If this is accomplished, one will be left with only the fluctuations
in 21cm line emission. 

During the epoch of reionization the 21cm signal originates from the
neutral hydrogen in the intergalactic medium.  HII regions ionized by
stars or quasars will appear as bubbles in the signal, creating a
``Swiss cheese'' topology.  On the other hand, the post-reionization
21cm signal is expected to originate from dense galactic regions which
are self-shielded from the ultraviolet background
\cite{postr1,postr2}.  Thus, the post-reionization signal is very
similar to other galactic emission lines.
Recently, statistical analysis of the post-reionization 21cm signal has
been attempted \cite{2010Natur.466..463C}.  Another idea related to
the present work is the cross correlation of the 21cm line with the
92cm line of deuterium \cite{deut}.

One important issue which does not arise in 21cm observations is
confusion from multiple lines.  With multiple lines of different rest
frame wavelengths the intensity at a particular
observed frequency corresponds to emission from multiple redshifts,
one for each emission line.  With both angle and frequency
information, the total emission corresponds to a superposition of 3D
maps of galaxies at different redshifts.

Fortunately, it is possible to statistically isolate the fluctuations
from a particular redshift by cross correlating the emission from two
different lines.  If one compares the fluctuations at two different
frequencies, which correspond to the same redshift in two different
emission lines, their fluctuations will be strongly correlated.
However, the signal from any other lines arises from galaxies at
different redshifts which are very far apart and thus will have much
weaker correlation.

We propose observations of these cross correlations, and show that
they can be described quantitatively by the cross power spectrum of
line emission.  This technique is particularly suitable for learning
about a large sample of faint sources which in a reasonable amount of
time can only be detected statistically.  As an illustrative example,
we calculate the signal-to-noise ratio of the line cross power
spectrum which could be measured by the proposed Space Infrared
Telescope for Cosmology and Astrophysics (SPICA) \cite{SPICAref}.  We
find that with SPICA, our technique can potentially probe line
emission from faint galaxies beyond a redshift $z=8$.  This would
contain information about the evolution of distant galaxies including
their metallicity, bias and duty cycle.  Such observations could also
potentially constrain the reionization history based on the minimum
mass of galaxies.  As another example, we consider cross correlating
CO line emission observed with the Cornell Caltech Atacama Telescope
(CCAT) \cite{CCATref} and 21cm radiation from galaxies measured by an
interferometer similar in scale to the Murchison Widefield Array (MWA)
\cite{MWAref}, but optimized for post-reionization redshifts.

The paper is organized as follows.  In $\S$2 we introduce and describe
the 3D line cross power spectrum technique.  In $\S$3 and $\S$4 we
discuss SPICA and CCAT plus our MWA-like interferometer respectively,
which we use to use illustrate the effectiveness of the proposed
technique.  Finally, we discuss and summarize our conclusions in
$\S$5.  Throughout, we assume a $\Lambda$CDM cosmology with
$\Omega_\Lambda=0.73$, $\Omega_m=0.27$, $\Omega_b=0.0456$, $h=0.7$,
and $\sigma_8=0.81$ \cite{wmap7}.

\section{Method}
\subsection{Line cross power spectrum}
We begin by introducing the cross power spectrum of line emission.  We
assume an observation that records both spatial and spectral data over
a patch of the sky.  We further assume that the spectrally smooth
foreground, including galaxy continuum emission can be subtracted
accurately.  This could be done by fitting a smooth function of
frequency to the data in different locations on the sky as has been
suggested for 21cm observations
\cite{2005ApJ...619..678M,Wang:2005zj,2006ApJ...653..815M,2009MNRAS.398..401L}.
Subtracting such a fit from the data would remove only the signal
which varies slowly as a function of frequency.  If we associate the
fluctuations with emission in a particular line, we can map each pixel
to a location in redshift space.

The fluctuation signal at a particular angle on the sky and
observed frequency, $\Delta
S(\theta_1,\theta_2,\nu)=S(\theta_1,\theta_2,\nu)-\bar{S}$, will have
several different components,
\begin{equation}
\label{deltas}
 \Delta S_1(\theta_1,\theta_2,\nu)=\Delta S_{\rm line1}+\Delta S_{\rm noise}+\Delta S_{\rm badline1}+\Delta S_{\rm badline2}+ \Delta S_{\rm badline3} + \ldots
\end{equation}
These include fluctuations in line emission from
galaxies at the target redshift in the associated line, detector noise,
and emission from galaxies at different redshifts in other lines.  We
term these ``bad lines''.

To estimate the flux amplitude of line emission fluctuations we assume
that galaxies trace the underlying cosmological density field.  If we
consider linear scales, it follows that the line
fluctuations due to galaxy clustering are given by $\Delta
S=\bar{S}\bar{b}\delta(\vec{r})$, where $\bar{S}$ is the average line
signal, $\bar{b}$ is the luminosity weighted average galaxy bias, and
$\delta(\vec{r})$ is the cosmological over-density at a location
$\vec{r}$ corresponding to the observed angle and frequency.  For the
moment we ignore Poisson fluctuations due to the discrete nature of
galaxies, which will be introduced later.  We describe our method for
estimating the average signal and bias of emission lines in the next
subsection.

\begin{figure*}
\includegraphics[width=3in]{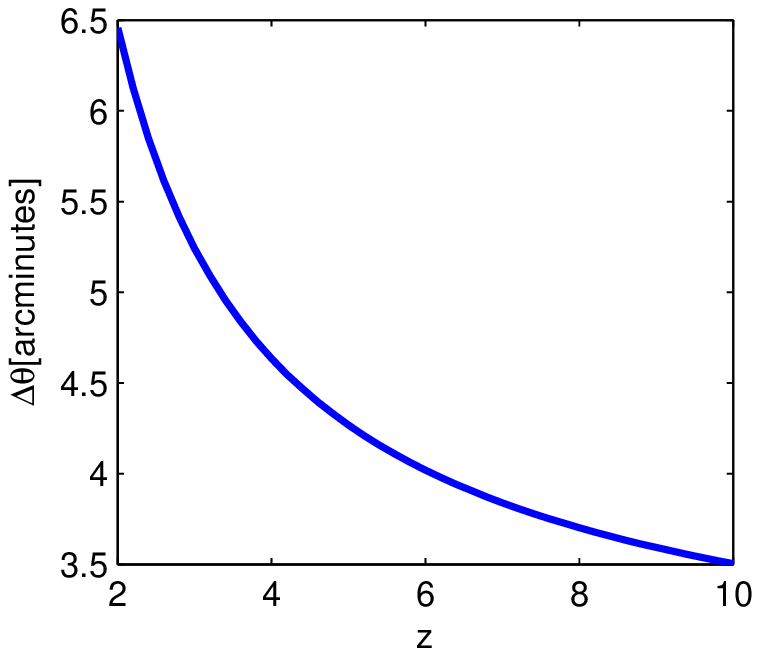}
\includegraphics[width=3in]{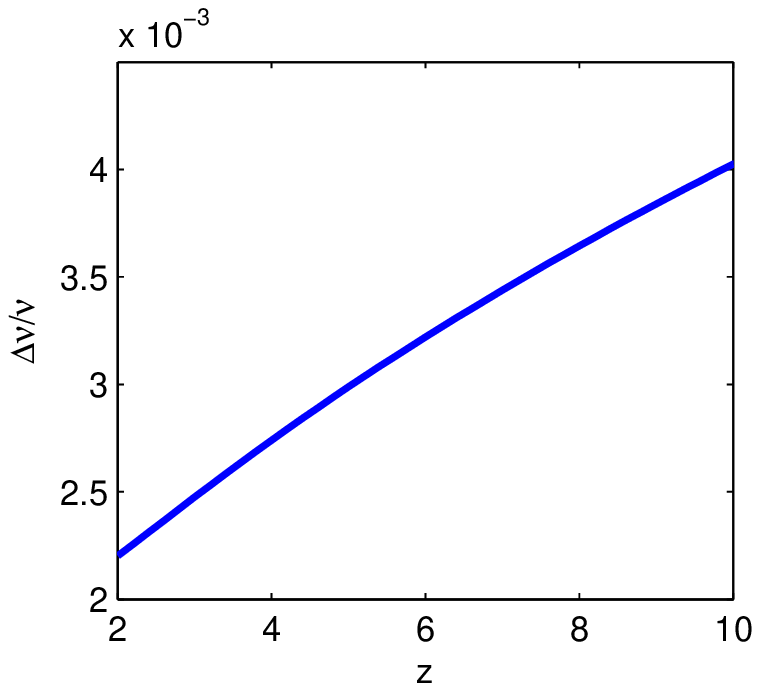}
\caption{\label{complot} The angle and frequency range corresponding
to 10 comoving Mpc. }
\end{figure*}

Ignoring peculiar velocities and using the flat sky approximation
(which applies for the small fields of view we consider in this
paper), we find,
\begin{equation}
\Delta S_{\rm line1}= \bar{S}_1 \bar{b} \delta (\vec{r}_o+\Delta \theta_1D_A \mathbf{\hat{i}} + \Delta \theta_2 D_A \mathbf{\hat{j}} + \Delta \nu \tilde{y}_1 \mathbf{\hat{k}}),
\end{equation}
where $\vec{r}_o$ is the comoving position corresponding to the
center of the survey, $\Delta \theta_1$ and $\Delta \theta_2$ are the
angular offsets from the center of the survey and $\Delta
\nu$ is the offset from the central observed frequency,
$D_A$ is the angular diameter distance in comoving units, and
$\tilde{y}$ is the derivative of comoving distance with respect to
observed frequency.  This last quantity is given by,
\begin{equation}
\tilde{y}= \frac{d\chi}{d\nu}=\frac{d\chi}{dz}\frac{dz}{d\nu}=\frac{\lambda_{r}(1+z)^2}{H(z)},
\end{equation}
where $\chi$ is the comoving distance to the observation, $\nu$ is the
observed frequency, $\lambda_{\rm r}$ is the rest frame wavelength of
a line, and $H(z)$ is the Hubble parameter.  Here $\mathbf{\hat{i}}$, $\mathbf{\hat{j}}$,
and $\mathbf{\hat{k}}$ are Cartesian unit vectors with $\mathbf{\hat{k}}$ pointing along
the line of sight of the survey.  In Figure \ref{complot} we show the
relation between angle, frequency, and comoving position as a function
of redshift.

Similarly, for the bad lines we have,
\begin{equation}
\Delta S_{\rm badline1}=\bar{B}_1 \bar{b}_{\rm 1} \delta (\vec{r}_o+ d_1\mathbf{\hat{k}}+\Delta \theta_1D_{A1} \mathbf{\hat{i}} + \Delta \theta_2 D_{A1} \mathbf{\hat{j}} + \Delta \nu \tilde{y}_{b1} \mathbf{\hat{k}}),
\end{equation}
where $d_1$ is the shift along the line of sight due to each bad line
being at a different redshift than the target line, $\bar{B}_1$ is the
average signal from the bad line, and $\bar{b}_1$ is the average bias of
the galaxies emitting in the bad line.  Note that the angular diameter
distance and derivative of comoving distance with respect to observed
frequency are evaluated at the redshift of the bad line galaxies which
we denoted with subscripts.

Instead of angle and frequency we label our observed pixels in terms
of the location in space corresponding to our target line
($\vec{r}_o,\vec{r}$), where $\vec{r}=x\mathbf{\hat{i}} + y\mathbf{\hat{j}} + z\mathbf{\hat{k}}$
is the distance from the center of the survey, with $x=\Delta \theta_1D_A$, 
$y=\Delta \theta_2 D_A$, and $z=\Delta \nu \tilde{y}$.
We rewrite Eq.~(\ref{deltas}) using these coordinates,
\begin{equation}
 \Delta S_1(\vec{r}_o,\vec{r})=  \Delta S_{\rm noise} + \bar{S}_1 \bar{b} \delta (\vec{r}_o+\vec{r}) + \bar{B}_1 \bar{b}_{\rm 1} \delta(\vec{r}_o+d_1\mathbf{\hat{k}} +\vec{r'}_1) + \bar{B}_2 \bar{b}_{\rm 2} \delta(\vec{r}_o+d_2\mathbf{\hat{k}} +\vec{r'}_2) + \ldots,
\end{equation}
where $\vec{r'}_n=c_{xn}x\mathbf{\hat{i}} +c_{yn}y\mathbf{\hat{j}} +c_{zn}z\mathbf{\hat{k}}$ for
the $n$'th bad line.  The constants $c_{xn},c_{yn}$ and $c_{zn}$
reflect the cosmological stretching of the data cube at different
redshifts, which are given by $c_{xn}=c_{yn}=D_{An}/D_A$ and $c_{zn}=\tilde{y}_{bn}/\tilde{y}_1$.

As mentioned in the Introduction, we can statistically remove the
 contribution from bad lines by cross correlating the fluctuations
 from two different lines originating from nearby locations.  We
 define the 2-point cross correlation function as,
\begin{equation}
\label{ccdef}
\xi_{1,2}(\vec{r})=\langle \Delta S_{1}(\vec{r}_o,\vec{x}) \Delta S_{2}(\vec{r}_o,\vec{r}+\vec{x}) \rangle,
\end{equation}
and the cross correlation power spectrum as its Fourier transform,
\begin{equation}
\label{cPSdef}
P_{1,2}(\vec{k}) = \int d^3\vec{r} \xi_{1,2}(\vec{r}) e^{i \vec{k} \cdot \vec{r}}.
\end{equation}

Note that the noise in different pixels will be uncorrelated and
that bad line emission from different lines will in general originate
from galaxies at different redshifts which will be essentially
uncorrelated.  Thus, we expect the cross correlation function and power
spectrum to depend only on galaxies at the target location,
\begin{equation}
\xi_{1,2}(\vec{r})=\langle \Delta S_{\rm line1}(\vec{r}_o,\vec{x}) \Delta S_{\rm line2}(\vec{r}_o,\vec{r}+\vec{x}) \rangle=\bar{S}_1 \bar{S}_2 \bar{b}^2 \langle \delta(\vec{x}) \delta(\vec{r}+\vec{x}) \rangle=\bar{S}_1 \bar{S}_2 \bar{b}^2 \xi(\vec{r}),
\end{equation}
where $\xi(\vec{r})$ is the cosmological matter correlation function.

\subsection{Average signal, average bias, and shot-noise}
To calculate the average signal and bias for a particular line, we
assume that the luminosity of a galaxy in a particular line is a
function of the mass of the halo hosting it: $L=L(M)$.  We also assume
that galaxies only emit a significant amount of radiation over some
fraction of cosmic time, the duty cycle, $\epsilon_{\rm duty}$. The
average signal is then (in erg/s/${\rm cm}^2$/Hz/Sr),
\begin{equation}
\label{avgsig}
\bar{S}=\int_{M_{\rm min}}^{\infty} dM\frac{L(M)}{4\pi D_L^2} \epsilon_{\rm duty}  \frac{dn}{dM}\tilde{y}D_A^2,
\end{equation}  
where $M_{\rm min}$ is the minimum mass of dark matter halos which can
host galaxies, and $dn(M)$ is the comoving density of halos of mass
between $M$ and $M+dM$.  Before reionization $M_{\rm min}$ is set by
the requirement that gas can cool efficiently via atomic hydrogen
cooling, corresponding to halos with virial temperatures greater than
$10^4\rm{K}$, whereas after reionization this is the threshold for
assembling heated gas out of the photo-ionized intergalactic medium,
corresponding to a minimum virial temperature of $10^5\rm{K}$
\cite{2008MNRAS.390.1071M,1992MNRAS.256P..43E,1996ApJ...465..608T,1997MNRAS.292...27H,Wyithe:2006st,1994ApJ...427...25S,2006ApJ...640....1B,1997ApJ...476..458H}.
$L(M)$ is the line luminosity (in ${\rm erg/s}$) from one dark matter
halo of mass $M$, $D_L$ is the cosmological luminosity distance, and
$\frac{dn}{dM}$ is the halo mass function in units of comoving density
per mass \cite{2001MNRAS.323....1S}.

Assuming that the flux in a line from a halo is proportional to the
mass of the halo, the average bias is given by,
\begin{equation} 
\label{avgbias}
\bar{b} = \frac{\int_{M_{\rm min}}^\infty \frac{dn}{dM}b(M,z)MdM}{\int_{M_{\rm min}}^\infty \frac{dn}{dM}MdM},
\end{equation}
where $b(M,z)$ is the bias associated with halos of mass $M$ at
redshift $z$ \cite{2001MNRAS.323....1S}.  We can calculate the average
signal and bias for both the target lines and bad lines using
Eqs.~(\ref{avgsig}) and (\ref{avgbias}).  Note that both of these
quantities must be calculated at the appropriate redshifts for the
different bad lines.

In addition to fluctuations due to the clustering of galaxies, there
are also Poisson fluctuations arising from the discrete nature of
galaxies which sample the cosmological density field.  These
fluctuations are unimportant on very large scales, but dominate on the
smallest scales.  Given our model of the line fluctuations, it is
straightforward to show from Eqs.~(\ref{ccdef}) and (\ref{cPSdef})
that the cross power spectrum of galaxy line emission from two lines
is given by,
\begin{equation}
 P_{1,2}(\vec{k})=\bar{S}_{1}\bar{S}_{2}\bar{b}^2 P(\vec{k}) + P_{\rm shot}, 
\end{equation}
where $P(\vec{k})$ is the cosmic power
spectrum of density fluctuations and the Poisson or shot-noise power spectrum due to
the discrete nature of galaxies is given by,
\begin{equation}
\label{shoteqn}
P_{\rm shot}=\int_{M_{min}}^{\infty} dM  \left (\frac{L_1(M)}{4\pi D_L^2} \right ) \left (\frac{L_2(M)}{4\pi D_L^2} \right ) \epsilon_{\rm duty}  \frac{dn}{dM}(\tilde{y}_1D_A^2)(\tilde{y}_2D_A^2),
\end{equation}
where the indices 1 and 2 denote lines 1 and 2.

\subsection{Cross power spectrum estimation}
Next we introduce an unbiased estimator of the cross power
spectrum and use it to derive an expression for the variance in
its measurement.  We begin by defining the Fourier amplitude of the
fluctuations,
\begin{equation}
\label{famp}
f_{\vec{k}} = \int d^3\vec{r} \Delta S(\vec{r_o},\vec{r})W(\vec{r})e^{i\vec{k} \cdot \vec{r}},
\end{equation}
where $W(\vec{r})$ is a window function which is constant over
the survey volume and zero outside the survey volume.  It is
normalized such that, $\int W d^3\vec{r}=1$.  The center of the survey
volume is denoted by $\vec{r}_o$.

The Fourier amplitude can be broken up into the different sources
of fluctuations, 
\begin{equation}
\label{fdec}
f^{(1)}_{\vec{k}}=f^{S1}_{\vec{k}} +f^{n1}_{\vec{k}} +f^{B1}_{\vec{k}} +f^{B2}_{\vec{k}} \ldots,
\end{equation}
the galaxy line fluctuations at the target redshift, detector noise, and each
of the bad lines coming from different redshifts.

Using the convolution theorem, we
rewrite the Fourier amplitude for the target line fluctuations as,
\begin{equation}
\label{conv}
f^{S1}_{\vec{k}}=\frac{1}{(2\pi)^3}\int d^3\vec{k'} \bar{S}_1 \bar{b} \delta(\vec{k'})W(\vec{k'}-\vec{k})e^{-i\vec{k'}\cdot\vec{r_o}}.
\end{equation}
A rectangular window function in real space has the k-space form,
\begin{equation}
W(k_x,k_y,k_z)=\frac{\sin(k_xa_x/2)}{(k_xa_x/2)} \frac{\sin(k_ya_y/2)}{(k_ya_y/2)} \frac{\sin(k_za_z/2)}{(k_za_z/2)},
\end{equation}
where $a_x$, $a_y$, and $a_z$ are the spatial dimensions of the survey along the $x$, $y$, and $z$ axes.
  
The Fourier amplitude defined above can now be used to estimate the
cross power spectrum of different lines. 
If we cross correlate the Fourier amplitude of two different lines
corresponding to the same location, $\langle
f^{(1)}_{\vec{k}}f^{(2)*}_{\vec{k}}\rangle$ , all terms except for the
target lines' fluctuations are greatly suppressed as discussed above.
Ignoring momentarily the fluctuations due to the discrete nature of
galaxies, this results in,
\begin{equation}
\langle f^{(1)}_{\vec{k}}f^{(2)*}_{\vec{k}}\rangle=\frac{1}{(2\pi)^6}\int\int d^3\vec{k'}d^3\vec{k''}
W(\vec{k'}-\vec{k})W(\vec{k''}-\vec{k})^*
\bar{S_{1}}\bar{S_{2}}\bar{b}^2\langle\delta(\vec{k'}) \delta(\vec{k''})^*\rangle e^{i(\vec{k''}-\vec{k'})\cdot \vec{r_o}}.
\end{equation} 
Note that the cosmological power spectrum is defined by,
\begin{equation}
\label{cosk}
\langle \delta(\vec{k'})\delta(\vec{k''})^* \rangle =
(2\pi)^3\delta^D(\vec{k'}-\vec{k''})P(k'),
\end{equation}
and that for a large survey $|W(\vec{k'}-\vec{k})|^2 \approx
(2\pi)^3\delta^D(\vec{k'}-\vec{k})/V$, where $\delta^D$ is a Dirac
delta function and $V$ is the volume of the survey.  With these substitutions
we find that,
\begin{equation}
\label{conv2}
\langle f^{(1)}_{\vec{k}}f^{(2)*}_{\vec{k}}\rangle=\bar{S}_{1}\bar{S}_{2}\bar{b}^2 P(\vec{k})/V,
\end{equation}
which is simply the clustering component of the cross power spectrum divided
by the volume of the survey.  When Poisson fluctuations due to the
discrete nature of galaxies are included it is straightforward to
show that this will give the total line cross power spectrum divided
by the volume of the survey.  Thus, we take our unbiased estimator to
be the real part of this quantity times the volume of the survey,
\begin{equation}
\hat{P}_{1,2}=\frac{V}{2} (f^{(1)}_{\vec{k}}f^{(2)*}_{\vec{k}}+f^{(1)*}_{\vec{k}}f^{(2)}_{\vec{k}} ).
\end{equation}

Equipped with an estimator, we can now calculate the variance on a
measurement of the cross power spectrum.  The error is given by,
\begin{equation}
  \label{plug}
\delta P_{\rm 1,2}^2 = \langle \hat{P}_{\rm 1,2}^2 \rangle - \langle \hat{P}_{\rm 1,2}\rangle^2.
\end{equation}
With this estimator we find (see appendix A),
\begin{equation}
\label{ccerror}
\delta P^2_{1,2}=\frac{1}{2}(P_{\rm 1,2}^2 + P_{\rm1total}P_{\rm2total}),
\end{equation}
where $P_{\rm1total}$ is the total power spectrum corresponding to the
first line being cross correlated,
\begin{equation}
\label{totpow}
P_{\rm 1total} = \bar{S}^2_{1}\bar{b}^2 P(\vec{k},z_{\rm target}) + P_{\rm noise} + \bar{B}^2_{1}\bar{b}_1^2 \frac{P(\vec{k'_1},z_1)}{(c_{x1}c_{y1}c_{z1})} +\bar{B}^2_{2}\bar{b}_2^2 \frac{P(\vec{k'_2},z_2)}{(c_{x2}c_{y2}c_{z2})}\ldots.
\end{equation}
Here we have the total power spectrum for each line, which includes
the line fluctuation power spectrum of the line at the target
redshift; a power spectrum due to the detector noise fluctuations; and
the power spectrum for each of the bad lines.  The $c$'s which appear
in the denominators of the bad line power spectra result from the fact
that the volume corresponding to the field of view on the sky and
frequency interval of the survey is different for the location of each
bad line.  Similarly, the wave-number for the various bad lines is changed to
$\vec{k'_1} =
\frac{k_x}{c_{x1}}\mathbf{\hat{i}}+\frac{k_y}{c_{y1}}\mathbf{\hat{j}} +
\frac{k_z}{c_{z1}}\mathbf{\hat{k}}$.  To avoid an overly cumbersome
equation, we have not explicitly written the shot-noise power spectra.
Note that for each clustering power spectrum which appears in this
equation there is a corresponding $P_{\rm shot}$ given by
Eq.~(\ref{shoteqn}).

Up to this point we have only been dealing with the error in an estimate
of one k-mode.  However, one can exploit the isotropy of the universe
and average the value of the cross power spectrum for a thin spherical
shell in k-space.  The error on the cross power spectrum changes as a
function of angle in k-space (due to $\vec{k'}$ appearing in
Eq.~\ref{totpow}).  One should take a weighted average, where modes in
the shell with a better signal-to-noise ratio are weighted more heavily.  The
optimal way to do this is an inverse variance weighted average of the
cross power spectrum in the shell.  It follows that
the error on the averaged cross power spectrum is given by,
\begin{equation}
\delta P(k)_{1,2} = \left (\sum_{\rm shell} \frac{1}{\delta P_{1,2}(\vec{k})^2} \right )^{-1/2},
\end{equation}
where we are summing over all of the modes in a shell.  Note that the
resolution of k-space is given by $2\pi/a_i$ in the $i$'th direction
for Cartesian coordinates.  We only add modes in the upper half-plane
because the power spectrum is the Fourier transform of a real-valued
function.  The maximum k-modes available are set by the angular and
frequency resolution of the observations, while the minimum values are
set by the dimensions of the survey volume.

\subsection{Multiple lines}
One could also combine the information from the cross correlations of
many lines.  For instance, one could do a weighted average of the cross
power spectra of all combinations of available lines,
\begin{equation}
P_{\rm AVG} = w_{1,2}\hat{P}_{1,2} + w_{1,3}\hat{P}_{1,3} + w_{2,3}\hat{P}_{2,3} \ldots
\end{equation}
where the $w$'s are weighting factors which would need to be highest
for the pairs of lines with the highest signal-to-noise ratio.  This
helps to detect the faintest galaxies.

\subsection{Remaining confusion}
As discussed above, if one cross correlates fluctuations from two
different lines at frequencies corresponding to a target redshift,
each will have its own set of bad lines from various other redshifts.
If a bad line from the first line originates from a redshift very
close to a bad line from the second line there may be a spurious
signal in the cross correlation of the fluctuations which is not from
the target redshift.

Fortunately, we do not expect this type spurious signal to be very
problematic.  If a pair of spurious bad lines is present one should be
able to cross correlate this pair directly (then the target lines will
appear as a spurious pair).  As long as the fluctuations from the
spurious pair are not much larger than the target lines, one will be
able to accurately subtract off their contribution.  In the examples
with SPICA considered below, we find that after all of the bright
sources which can be individually detected are removed, the
fluctuations from each bad line are smaller or comparable to the
fluctuations in the lines we cross correlate.

A factor which will aid in subtracting off this spurious signal is
that (ignoring redshift space distortions), the cross power spectrum
from the target lines will only depend on the
magnitude of $\vec{k}$ and not the direction.  However, the signal
from the spurious pair will change with the direction of the k-mode
due to the stretching of the data cube (i.e. they have different
$c_x$, $c_y$, and $c_z$ values defined above).

There is an additional effect which will further eliminate this
problem.  Two problematic bad lines will never originate from exactly
the same redshift.  Since they are from slightly different redshifts,
the 3D data cube corresponding to the location of the emitting
galaxies will be stretched by different amounts.  That means when we
cross correlate the same $\vec{k}$ in the target lines they will have
slightly different $\vec{k}$ values in each of the problematic bad
lines.  Since k-modes which have different $\vec{k}$ are uncorrelated,
this will serve to reduce the spurious signal.

\section{SPICA}
\subsection{Instrument}
We use the proposed Space Infrared Telescope for Cosmology and
 Astrophysics (SPICA) \cite{SPICAref} as an example of an instrument
 which could be used to cross correlate line emission as discussed
 above.  SPICA is a 3.5 meter space-borne infrared telescope planned
 for launch in 2017.  It will be cooled below $5$K, providing
 measurements which are orders of magnitude more sensitive than those
 from current instruments.  We also consider the Atacama Large
 Millimeter Array (ALMA) \cite{ALMAref}, but find that it is not well
 suited for the cross correlations discussed in this paper.  Due to the
small field of view and high angular resolution of this instrument, after
the detectable galaxies have been removed, we find that remaining statistical
signal cannot be detected above the detector noise.

We focus on SPICA's proposed high performance spectrometer $\mu$-spec
(H. Moseley, private communication 2009).  This instrument will
provide background limited sensitivity with wavelength coverage from
$~250-700 \mu m$.  A number of $\mu$-spec units will be combined to
record both angle and spectral data in each pointing, which will be
perfectly suited for the cross correlation technique described above.
We assume that spectra for $100$ diffraction limited pixels can be
measured simultaneously with a resolving power of $R=\nu/\Delta
\nu=1000$, this represents an optimistic design.  We expect that most of the signal contributed by
galaxies which cannot be observed directly will originate from lines too
narrow to be resolved with this resolution.  We have checked this by
associating halos of a given mass with a corresponding rotational
velocity. 

 Since our observations are background limited, the noise fluctuations
 will be dominated by shot-noise from the finite number photons
 originating from the smooth foregrounds which we assume have been
 subtracted off.  In the wavelength range considered, the dominant
 foregrounds are dust emission from the Milky Way galaxy and zodiacal
 light from the solar system.  We use COBE FIRAS data to estimate the
 brightness of the total foregrounds in the faintest $10\%$ region of
 the sky \cite{1998ApJ...508..123F}.  If the observation has pixels
 with solid angle $\Delta \Omega$ and frequency width $\Delta \nu$,
 the foreground flux fluctuations (in ${\rm erg/s/cm^2/Sr/Hz}$) in one
 pixel will be,
\begin{equation}
\label{nfluc}
\Delta S_{\rm noise} = \frac{E_\gamma}{At \Delta \Omega \Delta \nu}
(N_\gamma-\bar{N}_\gamma),
\end{equation}
where $E_\gamma$ is the energy per photon coming from the foreground,
$A$ is the area of the primary dish of the telescope, $t$ is the
integration time, $N_\gamma$ is the
number of photons gathered in a pixel during the integration time, and
$\bar{N}_\gamma$ is the average number of photons that are collected
from one pixel during the integration time.  From Poisson statistics
it follows that the variance due to the foregrounds in a pixel is 
given by,
\begin{equation}
\label{noisesig}
\sigma^2_{\rm noise} = \frac{\bar{S}_{\rm fg} E_\gamma}{At \Delta \Omega \Delta \nu},
\end{equation}
where $\bar{S}_{\rm fg}$ is the average surface brightness from the
foregrounds (in ${\rm erg/s/cm^2/Sr/Hz}$).  One can then derive (see
appendix A) the noise power spectrum,
\begin{equation}
P_{\rm noise} = \bar{S}_{\rm fg1} \frac{E_\gamma D_A^2\tilde{y}}{At}.
\end{equation}

In deriving this equation we have implicitly assumed that the arrival
of individual photons are statistically independent from one another.
This will be true for foreground radiation at the wavelengths we
consider with SPICA.  However, at wavelengths longer than $\sim 1$mm
this assumption might no longer apply.  In general, one needs to use
the full equation for photon noise which includes correlations from
photons arriving in the same quantum state.  This correlation in the
arrival time of photons, so called ``photon bunching'', will increase
the noise calculated with simple Poisson statistics for individual
photons \cite{2003ApOpt..42.4989Z,2008JCAP...04..021L}.

\subsection{Relevant lines}
Atomic and ionic fine-structure lines in the far infrared are very
important in cooling the gas in galaxies.  They produce bright
emission lines which can be seen in distant sources.  For galaxies
with redshifts of $z \gtrsim 5$ many of these lines will be observed
within SPICA's wavelength range.

In order to estimate the amplitude of line emission fluctuations we
assume a linear relationship between line luminosity, $L$, and star
formation rate, $\dot{M}_*$.  The line luminosity from a galaxy is
then given by $L=\dot{M}_* \cdot R$, where $R$ is the ratio between
star formation rate and line luminosity for a particular line.  This
is similar to existing relations in different bands (see
\cite{Kennicutt}) and was used in the past to estimate the strength of
the galactic lines \cite{2008A&A...489..489R}.  For the first 7 lines
in Table \ref{tab}, we use the same ratios, $R$, as
\cite{2008A&A...489..489R} which were calculated by taking the
geometric average of the ratios from an observational sample of lower
redshift galaxies \cite{2001ApJ...561..766M}.

We also list line strengths for CO lines which could be observed with
other instruments (see $\S$4).  For transitions below CO(8-7), we use
the same $R$ values as \cite{2008A&A...489..489R}, which are
calibrated from observations of the galaxy M82
\cite{2005A&A...438..533W}.  For the higher CO transitions we
calibrate with the observations of M82 presented in Ref.
\cite{2010A&A...518L..37P}.  The $R$ values of M82 are representative
of CO emission from high redshift galaxies (see Figure 1 in
Ref. \cite{2008A&A...489..489R}).

We calculate the star formation rate in a halo by denoting the
fraction of gas in a halo which forms stars as the star formation
efficiency, $f_*$.  We then approximate the star formation rate as
constant over the duty cycle of the galaxy.  Following
(\cite{Wyithe:2006st,2005ApJ...620..553L,2007ApJ...668..627S}) we
write,
\begin{equation}
\dot{M_*}(M) = \frac{f_* (\Omega_b / \Omega_m) M}{\epsilon_{duty} t_H},
\end{equation}
where $t_H=0.97[(1+z)/7]^{-3/2}$Gyr is the age of the Universe at the 
high redshifts of interest.  The line luminosity of a galaxy is
then given by,
\begin{equation}
L=6.6\cdot 10^{6} \left (\frac{R}{3.8\cdot 10^{6}} \right) \left (\frac{M}{10^{10}M_\odot} \right) \left (\frac{1+z}{7} \right)^{3/2}\frac{f_*}{\epsilon_{\rm duty}}L_{\odot}.
\end{equation}

\begin{table}
\caption{\label{tab}Assumed ratio between star formation rate,
$\dot{M}_*$, and line luminosity, $L$, for various lines.  For the
first 7 lines this ratio is measured from a sample of low redshift
galaxies.  The other lines have been calibrated based the galaxy M82.  We obtain
the luminosity in a line from: $L[L_{\odot}]=R \cdot
\dot{M}_*[M_\odot~{\rm yr}^{-1}]$. }
\begin{tabular}{|c|c|c|}
\hline
Species & Emission Wavelength[$\mu$m] & R[$L_{\odot}/(M_{\odot}/yr)$] \\
\hline
CII & 158 & $6.0 \times 10^6$\\
OI  & 145 & $3.3 \times 10^5$\\
NII & 122 & $7.9 \times 10^5$\\
OIII & 88 & $2.3 \times 10^6$\\
OI & 63 & $3.8 \times 10^6$\\
NIII & 57 & $2.4 \times 10^6$\\
OIII & 52 & $3.0 \times 10^6$\\
CO(1-0)  &2610 & $ 3.7 \times 10^3$\\
CO(2-1)  &1300 & $ 2.8 \times 10^4$\\
CO(3-2)  &866 & $  7.0 \times 10^4$\\
CO(4-3)  &651 & $  9.7 \times 10^4$\\
CO(5-4)  &521 & $  9.6 \times 10^4$\\
CO(6-5)  &434 & $  9.5 \times 10^4$\\
CO(7-6)  &372 & $  8.9 \times 10^4$\\
CO(8-7)  &325 & $  7.7 \times 10^4$\\
CO(9-8)  &289 & $  6.9 \times 10^4$\\
CO(10-9) &260 & $  5.3 \times 10^4$\\
CO(11-10) &237  & $ 3.8 \times 10^4$\\
CO(12-11) &217  & $ 2.6 \times 10^4$\\
CO(13-12) &200  & $ 1.4 \times 10^4$\\
CI  & 610 & $ 1.4 \times 10^4$\\
CI  & 371 & $ 4.8 \times 10^4$\\
NII & 205 & $ 2.5    \times 10^5$\\

\hline
\end{tabular}
\end{table}

\begin{figure*}
\includegraphics[width=3in]{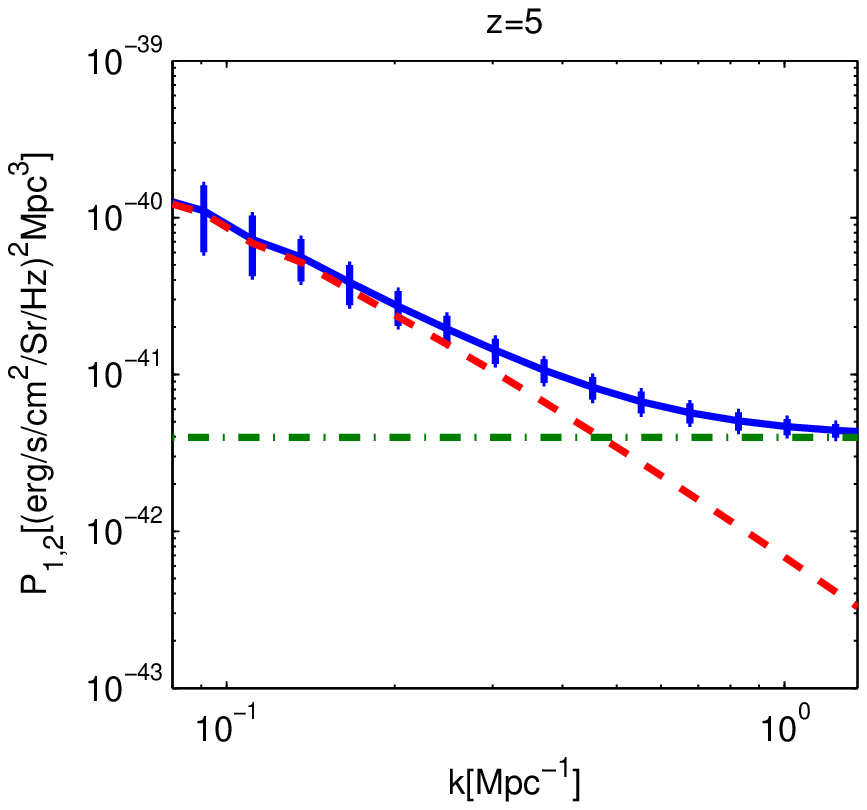}
\includegraphics[width=3in]{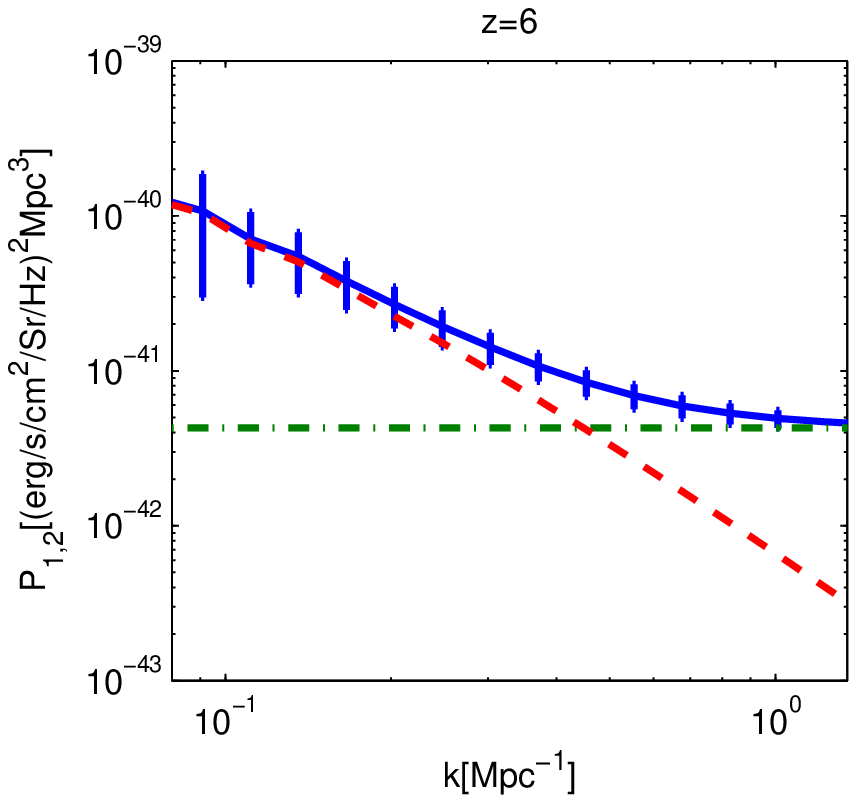}
\includegraphics[width=3in]{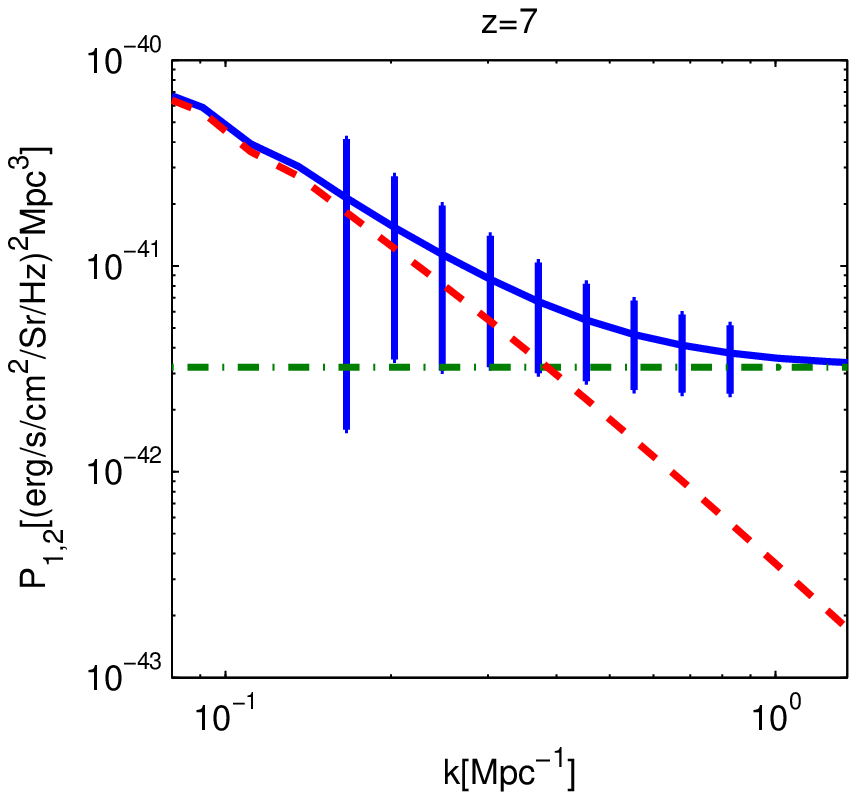}
\includegraphics[width=3in]{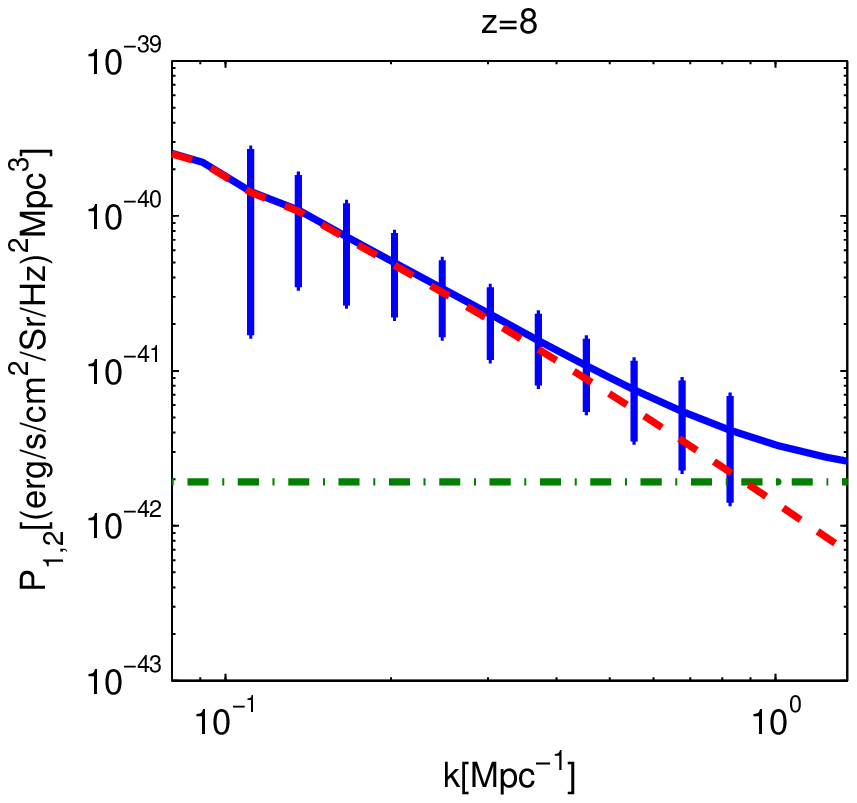}
\caption{\label{SNplots} The cross power spectrum of the OI(63$\mu$m)
and OIII(52$\mu$m) lines at redshifts of $z=5-8$.  Bright sources
which can be individually detected have been removed.  The dashed line
is the clustering component, the dot-dashed line is the shot-noise
power spectrum due to the discrete nature of galaxies and the solid
line is their sum.  The error bars show the root-mean-square error on
the cross power spectrum for a $10^6$ second observation with SPICA.
We assume that the survey is comprised of 256 adjacent pointings on the
sky and has a depth of $\Delta z = 0.1(1+z)$.
In the $z=5-7$ panels we have assumed that reionization occurred at a
much higher redshift.  In the $z=8$ panel we have assumed that
reionization occurred instantaneously at $z=6$; this increases the
signal to noise because it reduces the minimum mass of galaxies.  The
errors have been determined by averaging the cross power spectrum in
bins of width $\Delta k = k/5$.}
\end{figure*}

\begin{figure*}
\includegraphics[width=3in]{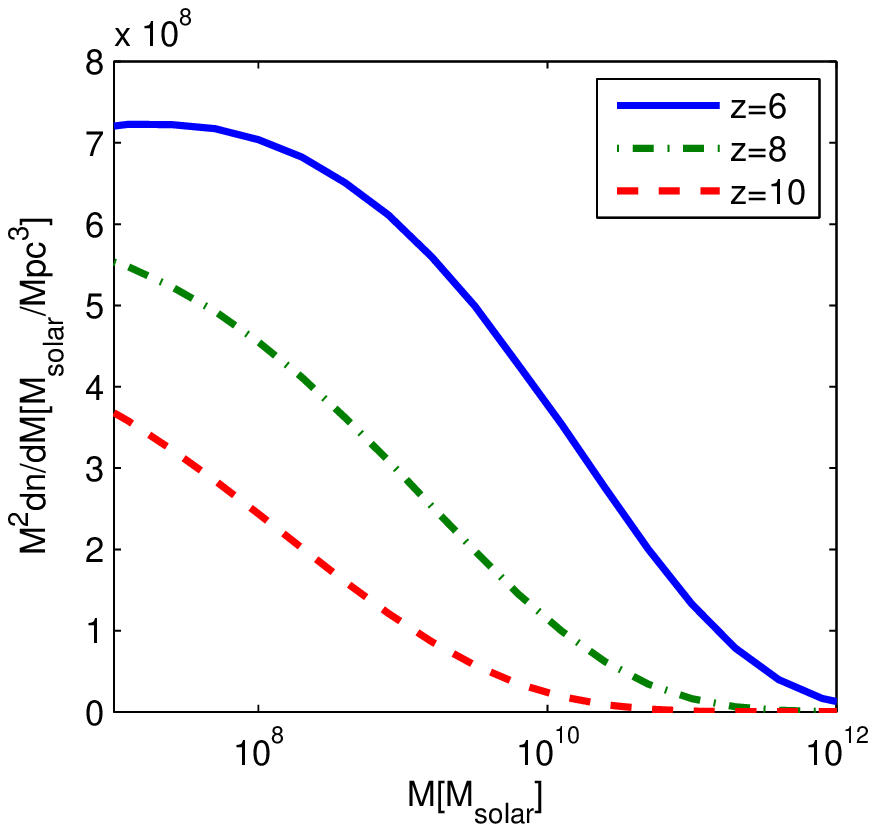}
\includegraphics[width=3in]{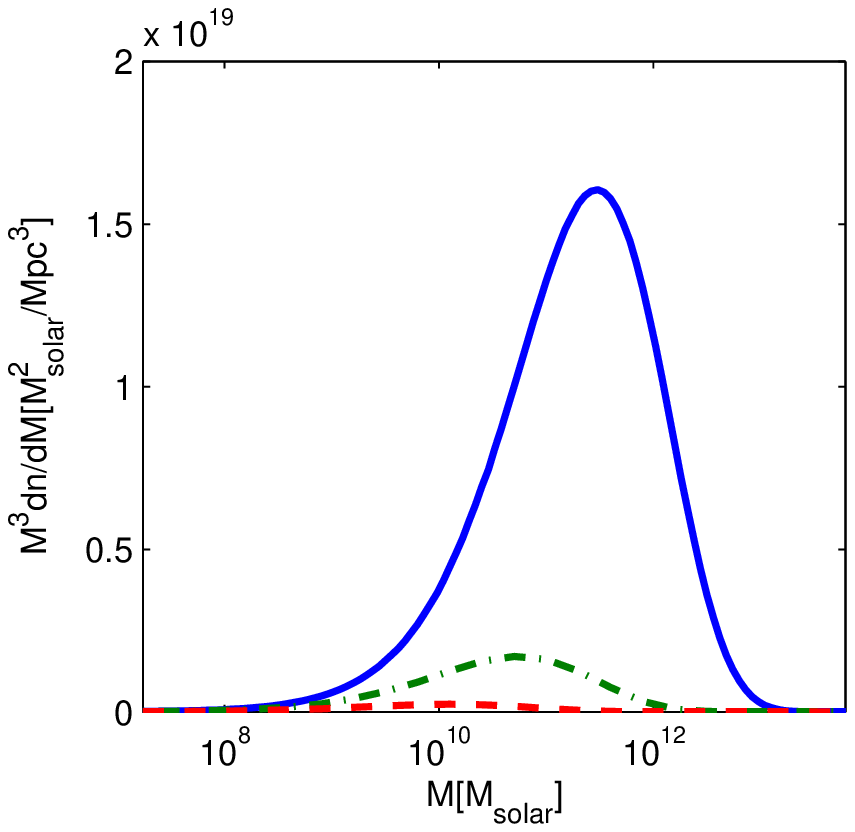}
\caption{\label{contplot} The mass distributions $M^2\frac{dn}{dM}$
(left) and $M^3\frac{dn}{dM}$ (right) versus $M$ at various redshifts.
The area under the first is proportional to the contribution to the
average line signal (see Eq.~(\ref{avgsig})) from the corresponding
mass range, while for the second it is proportional to the
contribution to the shot-noise power spectrum (see
Eq.~(\ref{shoteqn})).  We see that most of the average signal and thus
the clustering component of the cross power spectrum comes from low
mass halos.}
\end{figure*}

\begin{figure*}
\includegraphics[width=3in]{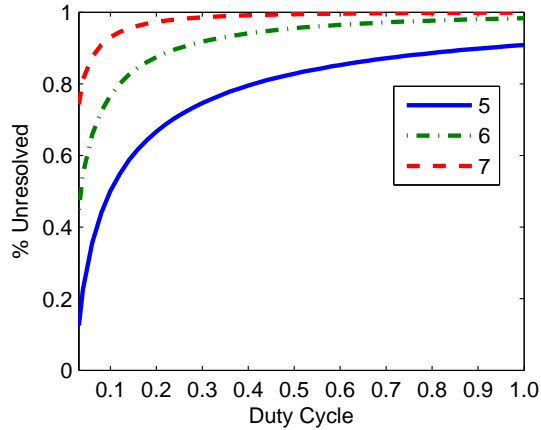}
\caption{\label{pduty} The percentage of the average
OI(63 $\mu$m) line signal which originates from galaxies that are
less than 3$\sigma$ peaks in the noise versus duty cycle.  We assume a 
$10^6$s observation with 256 different pointings.  As the duty
cycle goes up there are more halos hosting fainter galaxies increasing the
amount of signal which will come from galaxies which cannot be directly 
detected.}
\end{figure*}

\begin{figure*}
\includegraphics[width=3in]{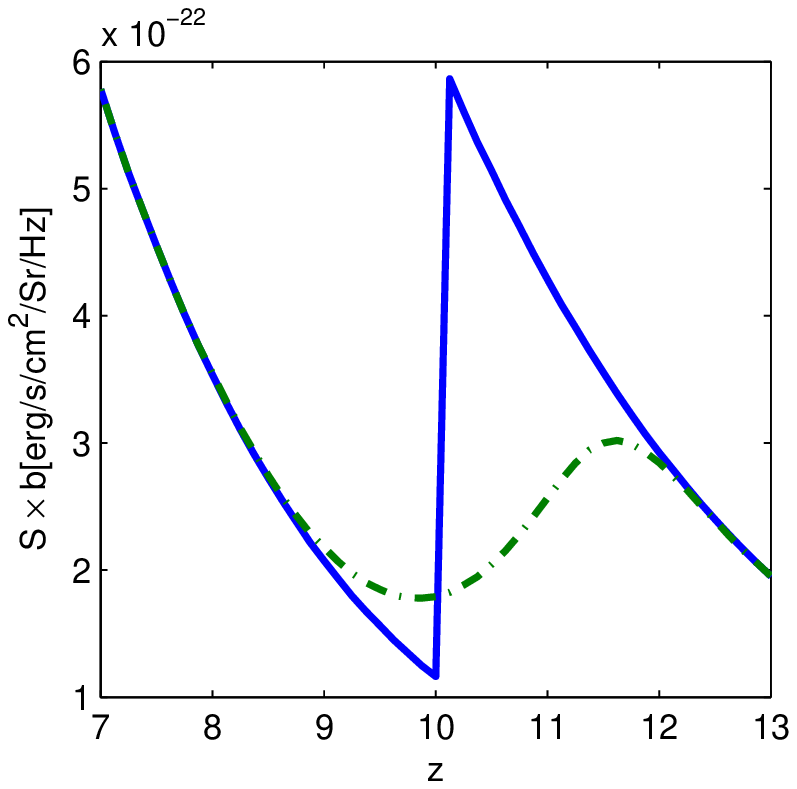}
\caption{\label{reionplot} The average OI(63 $\mu$m) line signal times
the bias as a function of redshift.  For the solid curve we have
assumed that reionization occurs instantaneously at $z=10$ and that
the minimum mass of galaxies changes from the requirement that
hydrogen cooling is efficient to the Jean's mass after reionization.
In the other curve, the change in minimum mass has been smoothed with
an error function.  While these may not be realistic reionization
histories, this figure illustrates that one may probe the reionization
history with the cross power spectrum. Though we have shown the cross
power spectrum corresponding to reionization at $z=10$, note that it
may be difficult to observe this signal with SPICA.}
\end{figure*}

\subsection{Results}
To demonstrate the effectiveness of the cross correlation technique,
we calculate the signal-to-noise ratio for the line cross power
spectrum of OI($63 \mu m$) and OIII($52 \mu m$) measured with SPICA.
Because this technique is most useful for detecting faint galaxies
which cannot be individually detected, we assume that all of the
pixels containing very bright line emission have been removed.  This
is done for both the lines we are cross correlating and the bad lines.
We assume that all pixels with line emission corresponding to
5$\sigma$ peaks when compared with the foreground noise are removed.
This effectively changes the upper limits of integration in
Eqs.~(\ref{avgsig}) and (\ref{avgbias}).  We find that this only
requires removing a small fraction of the available pixels.  All lines
in Table \ref{tab} are used as bad lines.

In order to calculate which dark matter halos are bright enough to be
removed, we assume that all of the emission from a halo appears in one
pixel.  We expect this to be a good approximation, since most of the
signal from high redshift galaxies originates in halos with virial
radii corresponding to angles smaller than the pixels of the
instrument we consider.  Additionally, sub-halos which constitute more
than $\approx 10\%$ of the parent halo are expected to sink to its
center by dynamical friction in less than a Hubble time
\cite{sink1,sink2}; this implies that each halo will most of the time
have one dominant galaxy at its center and only much fainter
satellites. 

There is another argument which justifies our one pixel assumption for
removing the detectable galaxies.  For low redshift galaxies
contributing to the bad line noise in the power spectrum, we find that
the minimum mass of halos which have galaxies that can be directly
detected and removed is smaller than halos which host more than one
galaxy.  The number of galaxies per halo above this mass scales
roughly linearly with halo mass \cite{halonumber}.  Thus, if the
galaxies are equally luminous in halos which host multiple galaxies,
all of them can be directly detected and removed.  When they are not
equally luminous most of the signal will originate from the brighter
galaxies which will be removed. 

In all of our calculations we use the linear power spectrum computed
with CAMB \cite{CAMBref}.  We expect the linear power spectrum to be a
good approximation.  The smallest scales probed in our examples are
still in the linear regime at the redshifts of interest.  On these scales
we may be in the nonlinear regime of the matter power spectra at lower
redshifts which appear in the bad line noise terms.  One
way to approximate the nonlinear power spectrum is to use the so
called Halo Model (see \cite{halomodel} and references therein).  In
this treatment, correlations are separated into two terms: the 1-halo term
corresponding to correlations between matter in the same halo and the
2-halo term corresponding to correlations between matter in different
halos.  After removing bright sources we expect the signal to be
coming from the center of halos.  Thus, the 1-halo contribution to our
cross power spectrum will originate only from the shot-noise power spectrum
given in Eq.~(\ref{shoteqn}).  The linear power spectrum should give a
reasonable approximation to the 2-halo term.

We adopt fiducial values of the duty cycle and star formation
efficiency of $\epsilon_{\rm duty}=0.1$ and $f_*=0.1$
\cite{2007ApJ...668..627S}.  The total observation time is assumed to
be $10^6$ seconds.  During this time we assume that 256 adjacent
pointings are taken to provide a mosaic over an area of the sky.
Increasing the number of pointings reduces the error in the power
spectrum due to sample variance, including noise provided by bad
lines, but increases the error due to the detector noise power
spectrum since fewer photons are collected in each pointing.
For the depth of the survey along the line of sight we assume
$\Delta z = 0.1(1+z)$.  We assume that the
cosmological evolution across this redshift range is negligible.

In Figure \ref{SNplots}, we plot the OI($63 \mu$m) and OIII($52 \mu$m)
cross power spectrum with error bars.  For the plot at $z=5$, 6 and 7
we have assumed that reionization occurred at a much higher redshift,
such that the minimum mass of halos which host galaxies is determined
from the Jeans mass in the photo-ionized IGM with a minimum virial
temperature of $10^5$K
\cite{2008MNRAS.390.1071M,1992MNRAS.256P..43E,1996ApJ...465..608T,1997MNRAS.292...27H,Wyithe:2006st,1994ApJ...427...25S}.
For the plot at $z=8$ we assumed that reionization occurred
instantaneously at $z=6$ so that the minimum mass of galaxies at the
target redshift is set by the condition that there is efficient atomic
hydrogen cooling with a minimum virial temperature of $10^4$K
\cite{2006ApJ...640....1B,1997ApJ...476..458H}.  Because this
corresponds to a smaller minimum mass there is more signal, improving
the signal-to-noise ratio.  In all of the plots we show the clustering
and shot-noise components which make up the power spectrum.  In all
cases the clustering signal dominates on large scales.  The error bars
are much smaller for high k-values because they represent clustering
on small scales and there are many more small scale regions to sample
within a given survey volume.

The OI($145 \mu$m) bad line for OI($63 \mu$m) and the NII($122 \mu$m)
bad line for OIII($52 \mu$m) originate from very nearby redshifts.
Thus, the cross correlation of these could produce a spurious signal.
However, we find that for our $z=5-6$ examples all of the galaxies
emitting this spurious signal could be located with the CII($158 \mu$m) line
and removed.  For the $z=7-8$ examples one may need to measure and
subtract away some of the spurious signal as discussed above.  This
was not considered for Figure \ref{SNplots} and could slightly degrade
the constraints on the cross power spectrum at these redshifts.

Figure \ref{contplot} shows $M^2\frac{dn}{dM}$ and $M^3\frac{dn}{dM}$
versus $M$.  The area under the first distribution is proportional to
the contribution to the average signal from the corresponding halo
mass range.  Similarly, for the second function the area under the
curve is proportional to the contribution to the shot-noise power
spectrum.  Fainter sources contribute more to the clustering signal
because of their great numbers.  By contrast, the shot-noise signal
originates mainly from bright sources.  If the brightest sources are
removed, then the shot-noise signal should be substantially reduced
while the clustering signal remains relatively unchanged.  Owing to
the hierarchical nature of structure formation in the universe more
matter will be located in smaller halos at higher redshifts.

Figure \ref{pduty} shows the fraction of the average OI(63 $\mu$m)
signal which originates from lines that would be detected at less than
$3\sigma$ significance for different values of the duty cycle.  The
average signal does not depend on the duty cycle because at lower
values, the brightness of galaxies in our model goes up by the same
factor that reduces the number of active galaxies.  Thus, our
technique is particularly effective if the duty cycle is high.  This
results in a higher number of fainter galaxies which would be harder
to detect directly, but just as easy statistically using the cross
power spectrum.

Figure \ref{reionplot} shows the average signal times the bias as a
function of redshift for different reionization histories.  We show
one line assuming that reionization is sudden and that $M_{\rm min}$
changes instantaneously and a line which smooths the transition of $M_{\rm
min}$ with an error function.  For the smoothed case, the minimum mass
of halos hosting galaxies is given by,
\begin{equation}
M_{\rm min}(z) = \left (M_1-\frac{(M_1-M_2)}{2} \left [1+{\rm erf}(z-z_r) \right ] \right ) \left ( \frac{10}{1+z} \right )^{3/2},
\end{equation}
where $M_1=3 \cdot 10^9 M_{\odot}$, $M_2=10^8 M_{\odot}$, and the
redshift of reionization $z_r=10$.  While these are not necessarily
realistic reionization histories, they illustrate that the minimum
mass of halos which host galaxies could be detected with the cross
correlation technique.  It may be difficult to observe the cross power
spectrum at $z=10$.  If reionization occurs later it will be easier to
study.  If many different lines are used one may be able to probe
higher redshifts than those shown in Figure \ref{SNplots}.

\section{Cross-correlating CO lines with 21-cm emission}
\subsection{Instruments}
As another example, we consider cross correlating CO line emission
measured with a large ground based telescope and 21-cm line emission
from neutral hydrogen in galaxies measured with an interferometer
optimized for frequencies corresponding to post-reionization
redshifts.  As discussed above, in order to measure fluctuations in
line emission, one must first fit and subtract away the bright
foreground signal.  Because they will have completely different
sources of foreground emission, cross correlating CO and 21-cm line
emission could eliminate any residual foregrounds if they are present
after this subtraction process.

The Cornell Caltech Atacama Telescope (CCAT) is a large sub-mm
telescope to be built at high altitude in the Atacama region of
northern Chile \cite{CCATref}.  It will be a 25 meter telescope with a
10 arcminute field of view.  We consider a hypothetical instrument
which takes spectra of 1000 adjacent diffraction limited beams
simultaneously.  We assume spectrometers with resolving power of
$R=\nu/\Delta \nu=1000$ and background limited sensitivity on CCAT.
Specifically, we use the numbers listed in Figure 6 of
Ref. \cite{CCATsen}.

For the 21cm experiment, we consider a hypothetical interferometer which
is similar to the Murchison Widefield Array (MWA) \cite{MWAref}, but
optimized to observe 21-cm emission at a redshift of z=3.  We assume
an array of 500 tiles each consisting of 16 dipoles with an effective
area of $A_{\rm e}=2.8 \rm{m}^2$ per tile.

\begin{figure*}
\includegraphics[width=3in]{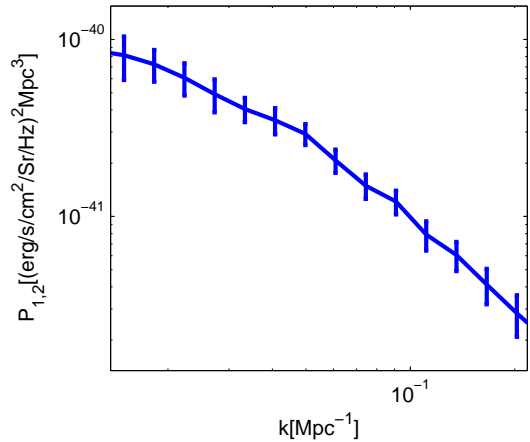}
\caption{\label{CCATfig} The cross power spectrum of CO(8-7) and 21-cm
galaxy line emission at $z=3$.  The shot-noise power spectrum is
negligible for the values of $k$ shown. The error bars show the
root-mean-square error on the cross power spectrum for a $10^6$ second
observation with CCAT and a separate $10^6$ second observation with
our hypothetical 21cm interferometer.  We assume a survey depth of
$\Delta z = 0.1(1+z)$.}
\end{figure*}

\subsection{Results}
To calculate the CO line signal we use the approach discussed in
$\S$3.2.  As in the SPICA example, we assume a duty cycle of
$\epsilon_{\rm duty}=0.1$ and star formation efficiency of $f_*=0.1$.
After reionization 21-cm emission is expected to come from
self-shielded neutral hydrogen in galaxies \cite{postr1,postr2}.
After reionization, the difference between the average observed 21-cm
brightness temperature from redshift $z$ and the CMB temperature today
is described by
\begin{equation}
\bar{T}_{\rm b} \approx 26\bar{x}_{\rm H} \left (
 \frac{\Omega_{\rm b}h^2}{0.022} \right)
\left(\frac{0.15}{\Omega_{\rm m}h^2} \frac{1+z}{10} \right)^{1/2} {\rm mK},
\end{equation}
where $\bar{x}_{\rm H}$ is the global mass-averaged neutral hydrogen
 fraction.  Observations have shown that out to $z \approx 4$ the
 cosmological density parameter of HI is $\Omega_{\rm HI} \approx
 10^{-3}$ \cite{Prochaska:2005wy}.  This corresponds to a
 mass-averaged neutral fraction of a few percent.  For our
 calculations we assume a value of $\bar{x}_{\rm H}=0.02$.  In the
 context of the formalism discussed above, $\bar{T}_{\rm b}$ is the
 average line signal after converting to units of
 $(\rm{ergs/s/cm^2/Hz/Sr})$ with the Rayleigh-Jeans Law.  With this
 quantity we can also calculate the shot-noise power spectrum as we
 did with Eq.~(\ref{shoteqn}).  For this calculation, we assume that
 the 21-cm flux originating from dark matter halos above $M_{\rm min}$
 is proportional to their mass.  The shot-noise power spectrum is
 lower than our previous example because we have emission from all
 halos above the minimum mass as opposed to a smaller fraction set by
 the duty cycle.

The noise power spectrum is given by
\begin{equation}
P_{\rm noise}(k \sin{\theta}) = D_{\rm A}^2 \tilde{y} \left(\frac{\lambda^2T_{\rm sys}}{A_{\rm e}}
\right)^2\frac{1}{t_0n(k \sin{\theta})},
\end{equation}
where $\lambda=21{\rm cm}\times (1+z)$ is the observed wavelength,
$T_{\rm sys}$ is the system temperature of the interferometer, $A_{\rm
e}$ is the effective area of each tile, $t_0$ is the total observing
time, and $n(k \sin{\theta})$ is the number density of baselines where
$k=|\vec{k}|$ and $\theta$ is the angle between $\vec{k}$ and the line
of sight.  The baseline density $n(k \sin{\theta})$ depends on the
geometrical configuration of the 500 antenna tiles.  We have assumed a
distribution of tiles with constant density within the innermost 8.9m
and which falls off as $r^{-2}$ out to 330m.  We assume that $T_{\rm
sys}=T_{\rm sky} + T_{\rm inst}$ where the sky temperature $T_{\rm
sky}=260 \left[ ({1+z})/{9.5} \right]^{2.55}$ \cite{Rogers:2008vh},
and that the instrumental temperature is $T_{\rm inst}=100$K.  For a
derivation of the expression which gives the noise power spectrum see
Ref. \cite{2006ApJ...653..815M}.

In Figure \ref{CCATfig} we plot the cross power spectrum of CO(8-7)
and 21-cm line emission at $z=3$.  We show the error bars corresponding
to an observation lasting $10^6$ seconds for each instrument and
covering the field of view of one primary beam,
$\Omega=\lambda^2/A_{\rm e}=0.25$ Sr, of our MWA-like interferometer.

\section{Discussion and conclusions}
In this paper, we have developed a formalism for measuring the cross
power spectrum of line emission from galaxies.  Cross correlating the
flux fluctuations in different lines from the same galaxies removes
the contaminating signal from other ``bad lines.''  We demonstrated
that it is possible to statistically measure the line fluctuation
signal from undetected galaxies at some target redshift.  The distinct
difference and advantage of this technique compared to traditional
galaxy surveys is that the signal originates from \emph{all} sources
of line emission rather than just the high signal-to-noise peaks in
the data.  In this way it is possible to study large populations of
faint galaxies in a reasonable amount of observing time.  Even though
individual galaxies will be difficult to detect, their large numbers
contribute significantly to the total signal.  The relative
contributions of different mass ranges to the clustering signal can be
seen in Figure \ref{contplot}.  At high redshifts, the faintest
galaxies produce the most signal per logarithmic mass interval.

The cross power spectrum of line emission measures the
product of the cosmic matter power spectrum, the average signal coming
from a pair of lines and the luminosity weighted bias of the source
galaxies.  For the standard set of cosmological parameters, one knows
the theoretical value of the matter power spectrum.  The cross power
spectrum then gives the product of the two average line signals with
the bias squared.  The shot-noise component of the cross
power spectrum depends on the duty cycle while the clustering
component does not.  This allows an estimation of the duty cycle by
comparing the cross power spectrum at low k-values where clustering dominates
to high k values where the shot-noise component dominates.

We demonstrated this technique by showing that atomic and ionic
fine structure lines from galaxies could be cross correlated using the
proposed SPICA mission.  We have found that for OI($63 \mu m$) and OIII($52
\mu m$) and an observation lasting a total of $10^6$ seconds, the
cross power spectrum of sources too faint to be directly detected can
be measured accurately out to $z \approx 8$.  This is important
because depending on the duty cycle, most of the line emission signal
originates from galaxies which cannot be detected otherwise.

The cross power spectrum at these redshifts allows one to measure the
total emission in O and N lines from a large sample of faint galaxies
as a function of redshift.  This information would constrain the
evolution of galaxy properties such as metallicity and star formation
rate.  It also constrains the bias and duty cycles of the source
galaxies.  Finally, the value of the average signal is sensitive to
the minimum mass of galaxies, and so the evolution in the average line
signal could constrain the reionization history.

We also consider an example where we cross correlate the CO and 21-cm
line emission from galaxies measured with CCAT and an interferometer
similar in scale to MWA, but optimized for post-reionization
redshifts.  As in the SPICA example, this would yield important
information about the evolution of CO emission from galaxies over
cosmic time.  Because we expect the foregrounds in CO and 21-cm
observations to be uncorrelated, cross correlation could eliminate
spurious residual foreground emission if any remained after the removal
process.  Of course, one could also cross correlate different
lines, such as the various CO lines and CII(158$\mu$m), with an
instrument like CCAT alone.

In our calculations we have assumed that the foregrounds are smooth in
frequency and can be removed perfectly, leaving only the fluctuations
due to line emission and detector noise.  If the foregrounds vary
rapidly as a function of angle on the sky it may only be possible to
measure k-modes along the line of sight.

The cross correlation technique is general and may be used for many
different lines (such as those listed in Table \ref{tab}).  It will be
most useful for instruments which have a large field of view, but
cannot detect individual faint sources effectively.  Additional
angular or spectral resolution will provide access to fluctuations on
smaller scales, but will not improve the accuracy of the cross power
spectrum on large scales.

\begin{acknowledgments}
We thank Mark Dijkstra and Jonathan Pritchard for useful comments.
We also thank the referee, Asantha Cooray, for helpful suggestions.
This work was supported in part by NSF grant AST-0907890 and NASA
grants NNX08AL43G and NNA09DB30A.
\end{acknowledgments}

\bibliography{paper}

\appendix
\section{Error on cross power spectrum}
The error on an estimate for the cross power spectrum at a
particular wave-vector is given by,
\begin{equation}
  \label{plug2}
\delta P_{1,2}^2 = \langle \hat{P}_{1,2}^2 \rangle - \langle \hat{P}_{1,2}\rangle^2 =  \left \langle \frac{V^2}{4} (f^{(1)}_{\vec{k}}f^{(2)*}_{\vec{k}} + f^{(1)*}_{\vec{k}}f^{(2)}_{\vec{k}})^2 \right \rangle  - P_{1,2}^2,
\end{equation}
where $V$ is the survey volume and $f_{\vec{k}}$ is the Fourier
amplitude defined in Eq.~(\ref{famp}) with superscripts denoting the
target lines being cross correlated.  For each target line we break up
$f_{\vec{k}}$ into terms corresponding to those in Eq.~(\ref{deltas}).
Plugging Eq.~(\ref{fdec}) into Eq.~(\ref{plug2}) and expanding we are
left with the sum of many products of four Fourier modes, many of
which are not correlated and vanish.  We are only left with products
of Fourier amplitudes that are correlated: detector noise with itself,
the bad lines' fluctuations with themselves and the target lines'
fluctuations with themselves and one another.  We then only need to
calculate the average value of these various products.

The terms involving detector noise are easily calculated by changing
the integral in Eq.~(\ref{famp}) into a sum.  It follows
that,
\begin{equation}
\langle Vf^{n1}_{\vec{k}} f^{n1*}_{\vec{k}}  \rangle = P_{\rm noise} = 
  \sigma^2_{\rm n} V_{\rm pix},
\end{equation}
where $\sigma^2_{\rm n}$ is the variance of the noise in units of
${\rm ergs/s/cm^2/Hz/Sr}$ in a single pixel and $V_{\rm pix}$ is the
survey volume corresponding to each pixel.

The Fourier amplitudes corresponding to the bad lines are given by
\begin{equation}
f^{B1}_{\vec{k}} =\bar{B}_1 \bar{b}_1 \int d^3\vec{r} \delta(\vec{r_o}+d_1\mathbf{\hat{k}}+\vec{r'}_1)W(\vec{r})e^{i\vec{k} \cdot \vec{r}},
\end{equation}
where $\vec{r'_1}=c_{x1}x\mathbf{\hat{i}} + c_{y1}y\mathbf{\hat{j}} + c_{z1}z\mathbf{\hat{k}}$.  We
change integration variables from $(x,y,z)$ to
$(x'_1=c_{x1}x,y'_1=c_{y1}y,z'_1=c_{z1}z)$, and proceed as we did in Eqs.
(\ref{conv})-(\ref{conv2}).  We find,
\begin{equation}
\label{blterm}
\langle V f^{B1}_{\vec{k}} f^{B1*}_{\vec{k}} \rangle =
\bar{B}^2_1\bar{b}_1^2 \frac{P(\vec{k'_1},z_1)}{(c_{x1}c_{y1}c_{z1})}, 
\end{equation}
where $\vec{k'_1} = \frac{k_x}{c_{x1}}\mathbf{\hat{i}}+\frac{k_y}{c_{y1}}\mathbf{\hat{j}} +
\frac{k_z}{c_{z1}}\mathbf{\hat{k}}$.  The $c's$ in these equations reflect the
stretching and squeezing of the data cube at the redshifts of the
spurious lines due to the redshift dependence of the angular diameter
distance and frequency per comoving interval as described above.

It is also necessary to calculate terms like,
\begin{multline}
V^2 \langle f^{S1*}_{\vec{k}} f^{S1*}_{\vec{k}}   f^{S2}_{\vec{k}} f^{S2}_{\vec{k}}  \rangle =\\ V^2
\frac{1}{(2\pi)^{12}}\int\int\int\int d^3\vec{k_1}d^3\vec{k_2}d^3\vec{k_3}d^3\vec{k_4}
W(\vec{k_1}-\vec{k})W(\vec{k_2}-\vec{k})^*
W(\vec{k_3}-\vec{k})^*  W(\vec{k_4}-\vec{k}) \\  \times \bar{S}_1^2\bar{S}_2^2\bar{b}^4
\langle \delta(\vec{k_1})\delta(\vec{k_2})^* \delta(\vec{k_3})^* 
\delta(\vec{k_4}) \rangle e^{-i(\vec{k_1}-\vec{k_2}-\vec{k_3}+\vec{k_4})\cdot 
\vec{r_o}}=2 \bar{S}_1^2 \bar{S}_2^2 \bar{b}^4 P(\vec{k}),
\end{multline}
where to calculate this integral we have used Wick's theorem $\langle
\delta_1\delta_2\delta_3\delta_4 \rangle=\langle \delta_1 \delta_2
\rangle \langle \delta_3 \delta_4 \rangle + \langle \delta_1 \delta_3
\rangle \langle \delta_2 \delta_4 \rangle +\langle \delta_1 \delta_4
\rangle \langle \delta_2 \delta_3 \rangle $.  We also used Eq.\
(\ref{cosk}) and the fact that for a large survey
$|W(\vec{k'}-\vec{k})|^2 \approx
(2\pi)^3\delta^D(\vec{k'}-\vec{k})/V$.

Putting all of this together we find,
\begin{equation}
\label{ccerror2}
\delta P^2_{1,2}=\frac{1}{2}(P_{1,2}^2 + P_{\rm1total}P_{\rm2total}),
\end{equation}
where $P_{\rm1total}$ is the total power spectrum corresponding to the
first line being cross correlated,
\begin{equation}
\label{totpow2}
P_{\rm 1total} = \bar{S}^2_{1}\bar{b}^2 P(\vec{k},z_{\rm target}) + P_{\rm noise1} + \bar{B}^2_{1}\bar{b}_1^2 \frac{P(\vec{k'_1},z_1)}{(c_{x1}c_{y1}c_{z1})} +\bar{B}^2_{2}\bar{b}_2^2 \frac{P(\vec{k'_2},z_2)}{(c_{x2}c_{y2}c_{z2})}\ldots
\end{equation}
The equation corresponding to the second line
being cross correlated is exactly the same, but of course contains the
noise at a different frequency and will have bad lines at different 
redshifts.  There will also be a shot-noise power spectrum for each clustering
spectrum included above, which was not written to avoid a cumbersome equation.

\end{document}